\documentclass[aip,reprint,citeautoscript]{revtex4-2}

\usepackage[colorlinks,urlcolor=blue,citecolor=blue]{hyperref}
\usepackage{color}
\usepackage[export]{adjustbox}
\usepackage{graphicx}
\usepackage{amsmath,amssymb}

\usepackage[per-mode=symbol]{siunitx}
\DeclareSIUnit\angstrom{\text{\AA}}
\DeclareSIUnit\atomicunit{\text{at.\,u.}}
\DeclareSIUnit\gpcc{\gram\per\centi\meter\cubed}

\begin{document}

\title{Nonlinear effects in light-ion stopping powers within real-time time-dependent density functional theory}

\author{Alina Kononov}
\affiliation{Center for Computing Research, Sandia National Laboratories, Albuquerque NM, USA \looseness=-1}

\author{Thomas W. Hentschel}
\affiliation{School of Applied \& Engineering Physics, Cornell University, Ithaca NY, 14850 USA \looseness=-1}

\author{Stephanie B. Hansen}
\affiliation{Pulsed Power Sciences Center, Sandia National Laboratories, Albuquerque NM, 87123 USA \looseness=-1}

\author{Andrew D. Baczewski}
\affiliation{Center for Computing Research, Sandia National Laboratories, Albuquerque NM, USA \looseness=-1}

\date{\today}

\begin{abstract}
Electronic stopping power models describing fuel heating processes in inertial fusion energy concepts typically assume linear-response behavior through quadratic scaling with the projectile charge.
We report the results of real-time time-dependent density functional theory (TDDFT) calculations indicating that even for low-Z ions, nonlinear processes modify stopping powers in warm dense matter by about 10\% near and below the Bragg peak.
By describing partial neutralization of slow ions, analytic effective charge models capture some qualitative aspects of the TDDFT results but do not always offer quantitative accuracy.
Cases where the effective charge inferred from TDDFT exceeds the bare ion charge suggest that more complex nonlinear effects also contribute.
These findings will inform future improvements to more efficient stopping power models.
\end{abstract}

\maketitle

\section{Introduction}
\label{sec:intro}

Electronic stopping power, the rate at which a charged particle loses energy to the electrons in a medium, drives fuel heating in fast-ignition fusion schemes \cite{roth2001fast} and leads to self-heating in inertial confinement fusion (ICF) \cite{zylstra_alpha-particle_2019}.
When the ion velocity $v$ is large compared to its nuclear charge $Z$, i.e., when $v \gg Z^{2/3}$ in atomic units, electrons cannot keep up with the moving charge, screening and electron capture processes become negligible, and linear-response theory remains valid \cite{sigmund:2008}.
However, ICF-relevant stopping powers regularly include lower-velocity regimes where these effects cannot be ruled out.
Although initial kinetic energies for proton driving beams or alpha fusion products --- typically on the order of \SI{10}{\mega\electronvolt} --- lie well within the linear-response limit, the entire velocity range is accessed as ions lose energy to the fusion target, and relatively low velocities near the Bragg peak at $v\sim1$ atomic unit (\SI{}{\atomicunit}) dominate energy deposition \cite{roth2001fast}.

Nonetheless, a wide range of stopping power models explicitly or implicitly assume linear response, which manifests as a quadratic dependence on $Z$ \cite{bethe:1930,fermi:1947,sigmund:2008,arista:1987,li1993charged,brown:2005,zylstra:2019}.
Intuitively, classical electrostatics predicts that a point charge induces a perturbation proportional to $Z$, leading to a $Z^2$ interaction energy between the charge and perturbation.
More formally, within linear-response theory, electronic stopping power is given by
\begin{equation}
    S_Z(v) = \frac{2Z^2}{\pi v^2} \int_0^\infty \frac{dk}{k} \int_0^{kv} d\omega \,\omega\, \mathrm{Im} \left[ \frac{-1}{\epsilon(k, \omega)}\right],
    \label{eq:lr}
\end{equation}
where $\epsilon(k, \omega)$ is the wave-vector- and frequency-dependent dielectric function of the medium \cite{lindhard:1964} and we use Hartree atomic units throughout unless otherwise noted.
Most practical stopping power models either assume a model dielectric function or effectively approximate the integrals of Eq.~\eqref{eq:lr} by assuming that certain types of excitations dominate in specific regimes.

In addition to the nearly ubiquitous linear-response assumption, efficient stopping power models often invoke other choices parameterizing the electronic response.
For example, earlier work has extensively scrutinized the treatment of collision frequencies \cite{hentschel:2023,hentschel2025statistical} and electronic screening in the target material \cite{Wang1998,Faussurier2010}.
The difficulty of focused experiments in the warm dense regime \cite{zylstra:2015,malko:2022} limits opportunities to constrain these input quantities, especially when they cannot be measured directly and their influence on observable information is convolved with other approximations.
Accurate calculations based on first principles offer alternative avenues for detailed benchmarking of more efficient models.

In this context, real-time time-dependent density functional theory (TDDFT) \cite{ullrich_time-dependent_2012,ullrich_time-dependent_2014} has emerged as a versatile benchmark-quality computational approach capable of predicting electronic response properties within linear response and beyond.
Recently, TDDFT calculations have informed choices made in more efficient linear-response and average-atom models of x-ray scattering \cite{hentschel:2023,baczewski:2021,hentschel2025statistical} and stopping power \cite{hentschel:2023,stanek_review_2024} in warm dense matter.
With an appropriately parameterized model dielectric function, the Lindhard formula of Eq.~\eqref{eq:lr} can successfully capture TDDFT predictions for proton-stopping in solids \cite{schleife:2015} and dense plasmas \cite{hentschel:2023}, particularly at projectile velocities beyond the Bragg peak.
However, stopping power models based on linear response depart from more accurate TDDFT results and empirical data \cite{ziegler:2010} (where available) for slow protons \cite{schleife:2015,hentschel:2023} and slow alpha particles \cite{stanek_review_2024} in degenerate matter.
These cases fall outside of the $v\gg Z^{2/3}$ regime, and the linear response assumption underlying Eq.\ \eqref{eq:lr} breaks down.

Slower and higher charge projectiles increasingly capture electrons from the medium into their bound states \cite{kononov:2020,kononov:2021,vazquez:2021,zhao:2015,Gauthier2013}, a process that cannot be described by the medium's dielectric response to a point charge alone.
Nonetheless, after a short time relative to the energy loss rate, the projectile's charge state is expected to equilibrate to a velocity-dependent average value that remains approximately constant over the time scale of further electronic excitations\cite{yost_electronic_2016,reeves2016electronic,lee_multiscale_2020,kononov:2020}.
Then, the linear response assumption might regain validity with an appropriately screened effective projectile charge, $Z \rightarrow Z_\mathrm{eff}(Z,v)$.
We review several proposed effective charge models in Section~\ref{sec:models}.

The accuracy of existing effective charge models and the extent to which they can improve stopping power predictions remain largely unexplored in the warm dense regime.
While experiments can measure ion charge states after transmission through a thin target, charge stripping processes upon exiting a material surface \cite{kononov:2020} preclude direct experimental access to effective charges even within ambient materials.
Challenges in producing and characterizing uniformly heated and/or compressed samples further compound the difficulty of validating these models in high-energy density systems.
Prior theoretical work has estimated equilibrium charge states in cold systems directly from the electron density simulated using TDDFT \cite{yost_electronic_2016,reeves2016electronic,lee_multiscale_2020,kononov:2020}, but difficulties in appropriately distinguishing electrons captured by the projectile, other screening charge within the projectile's wake, and the surrounding excited host frustrate efforts to relate such computed charge states to stopping powers.

Here, we investigate nonlinear effects influencing low-Z stopping powers in warm dense matter by comparing TDDFT results for different projectile charges.
In addition to protons and alpha particles, we consider artificial fractionally charged projectiles that enable direct access to effective charges and the linear-response limit.
Using the TDDFT data, we benchmark various analytic effective charge models for capturing deviations from ideal $Z^2$ scaling.
We also assess the sensitivity of the effective charge to target conditions and composition
through several cases: solid-density aluminum isochorically heated to an electronic temperature of \SI{1}{\electronvolt} or \SI{10}{\electronvolt} and carbon in thermal equilibrium at \SI{1}{\gram\per\centi\meter^3}, \SI{2}{\electronvolt} or \SI{10}{\gram\per\centi\meter^3}, \SI{2}{\electronvolt}.
We describe the TDDFT approach in Section~\ref{sec:methods}, present our computational results in Section~\ref{sec:results}, and summarize our findings in Section~\ref{sec:conclusions}.

\section{Effective charge models}
\label{sec:models}

The earliest effective charge models \cite{lamb:1940,bohr:1941,bohr:1948} derive from the simple assumption that captured electrons can only occupy bound states with orbital velocities (or binding energies) exceeding the projectile velocity (or kinetic energy).
One model of this type is the Bohr stripping criterion \cite{bohr:1948},
\begin{equation}
    Z_\mathrm{Bohr} = Z^{1/3} v,
    \label{eq:bohr}
\end{equation}
where validity is limited to $v< Z^{2/3}$ so that $Z_\mathrm{eff} \leq Z$.
Later work \cite{northcliffe1960energy,pierce1968stopping,brown1972stopping,betz1972charge} 
suggested modified forms like
\begin{equation}
    Z_\mathrm{Bohr}' = Z \left(1- e^{-Z_\mathrm{Bohr}/Z}\right) =  Z \left( 1 - e^{-v / Z^{2/3}} \right),
    \label{eq:bohrmod}
\end{equation}
which reduces to $Z_\mathrm{Bohr}$ in the low-velocity limit but recovers the correct high-velocity limit of $Z_\mathrm{eff} \rightarrow Z$ through exponential interpolation.
Variations on Eq.~\eqref{eq:bohrmod} include a scale factor in the exponent to control the transition between limiting regimes or other empirical parameters to improve agreement with experimental data.

Notably, the effective charge models in Eqs.~\eqref{eq:bohr}\,--\,\eqref{eq:bohrmod} depend only on projectile parameters.
Intuitively, thermal effects and material properties of the host system should also influence dynamic screening processes, including the projectile's ability to capture and retain electrons.
Gus'kov et al.\ \cite{gus2009method} thus proposed inserting the root-mean-square projectile velocity relative to host electrons, $v^*$, into an empirically parameterized version of Eq.~\eqref{eq:bohrmod} and obtained the following analytic approximation for a Fermi gas:
\begin{align}\begin{split}
    &Z_\text{Gus'kov} = Z\left( 1 - e^{-0.92 v^* / Z^{2/3} }\right), \\
    &(v^*)^2 = v^2 + 
    \frac{6}{5} \left( E_F + k_BT\,\frac{5\pi^2 /12}{\frac{E_F}{k_BT} +\pi^2/6} \right),
    \label{eq:guskov}
\end{split}\end{align}
where $E_F=\frac{1}{2}(3\pi^2 n_e)^{2/3}$ is the Fermi energy corresponding to electron density $n_e$ and $k_B$ is the Boltzmann constant.
Eq.~\eqref{eq:guskov} reduces to the form of Eq.~\eqref{eq:bohrmod} for low temperatures and densities, i.e., in the cold gas regime.

More sophisticated models go beyond the assumption of a point charge and account for the captured electron distribution $\rho$ by including a factor of $(Z-\rho(k))^2$ within the $k$ integral of Eq.~\eqref{eq:lr} \cite{brandt1982effective,clauser2018stopping}.
These approaches then invoke an approximate form for $\rho(k)$.
Here, we will focus on assessing the accuracy of the point charge models given by Eqs.~\eqref{eq:bohr}\,--\,\eqref{eq:guskov} as the simplest possible nonlinear correction to efficient stopping power models.

\section{Computational Methods}
\label{sec:methods}

Our approach closely follows the methodology described in Ref.\ \onlinecite{kononov2023trajectory}.
The electronic response of each host material to charged-particle irradiation was simulated using real-time time-dependent density functional theory (TDDFT) \cite{ullrich_time-dependent_2012,ullrich_time-dependent_2014} as implemented in a custom extension \cite{baczewski:2014,baczewski:2016,magyar:2016} of the Vienna \emph{ab initio} simulation package (\textsc{VASP}) \cite{kresse:1996a,kresse:1996b,kresse:1999}.
For aluminum, we considered only the free-electron contribution to stopping powers and thus explicitly treated only 3 valence electrons per ion through the projector augmented wave (PAW) method \cite{blochl:1994}.
For carbon, we performed both all-electron (AE) calculations and PAW calculations pseudizing the 1s states.
Exchange and correlation were treated with the adiabatic local density approximation (ALDA) \cite{zangwill:1980,zangwill:1981}, and large supercells allowed reciprocal-space sampling using the $\Gamma$ point only.

The electronic system of the host material began in the Mermin-Kohn-Sham equilibrium state \cite{mermin1965thermal}.
The aluminum ions were arranged in a cubic supercell with 256 atoms occupying face-centered cubic lattice sites, representing isochorically heated material.
Meanwhile, a single 125-atom carbon configuration at each density was drawn from an equilibrated \emph{ab initio} molecular dynamics simulation as described in Ref.~\onlinecite{melton2024transport}.
For each host configuration, the projectile's trajectory was optimized to representatively sample the simulation cell while avoiding artificial interaction with its own wake according to the strategies discussed in Refs.~\onlinecite{kononov2023trajectory,kononov2024reproducibility}.
For aluminum, we used the same trajectory analyzed in Ref.~\onlinecite{kononov2023trajectory}, which reported an 8\% finite-size error in proton stopping in ambient aluminum when compared to the most converged 500-atom setup considered in that work.
For carbon at \SI{10}{\gpcc} and \SI{1}{\electronvolt}, earlier work \cite{hentschel:2023} found a 5\% difference between proton-stopping powers computed with a larger 512-atom cell and a 128-atom cell comparable to those used in this work.
Since the stopping power shows only modest temperature dependence near these conditions, we expect similar convergence for the elevated temperatures considered here.

At an electronic temperature of \SI{1}{\electronvolt}, 5 electronic bands per aluminum ion sufficed to capture all Kohn-Sham orbitals with occupations of at least $3\times 10^{-6}$.
All other numerical parameters for the aluminum case match those previously used in Ref.\ \onlinecite{kononov2023trajectory}, which demonstrated excellent convergence with respect to plane-wave cutoff energy, time step, and reciprocal-space sampling for proton stopping in ambient aluminum.
Additional convergence tests for a smaller, 32-atom aluminum configuration revealed somewhat larger relative convergence errors with respect to time step for slow alpha particles (at most 14\% at $v=\SI{0.5}{\atomicunit}$ compared to 2.5\% for protons at the same velocity), and we include this source of error in our analysis of effective charges in aluminum.~\footnote{The computational cost of further converging the aluminum calculations was deemed too great --- requiring $\sim$0.5 machine-day on the Crossroads supercomputer at Los Alamos National Laboratory.}

Meanwhile, the AE (PAW) calculations for carbon at \SI{10}{\gram\per\centi\meter\cubed} and \SI{2}{\electronvolt} used 4.6 (3.8) bands per ion, capturing Kohn-Sham states with occupations of $4\times 10^{-6}$ ($1\times 10^{-6}$) or more.
The AE calculations for carbon at \SI{1}{\gram\per\centi\meter\cubed} and \SI{2}{\electronvolt} used 9.2 bands per ion and included Kohn-Sham occupations of $6\times 10^{-6}$ or more.
A plane-wave cutoff energy of \SI{2000}{\electronvolt} (\SI{1000}{\electronvolt}) sufficed to converge stopping powers within 1\% (0.5\%) relative to a \SI{3000}{\electronvolt} (\SI{2000}{\electronvolt}) cutoff in the AE (PAW) carbon simulations, as determined for the case of an alpha particle with $v=\SI{5}{\atomicunit}$.
The numerical time step varied with velocity such that the projectile traveled about $\SI{0.01}{\angstrom}$ ($\SI{0.02}{\angstrom}$) in each step of AE (PAW) carbon calculations.
Slow projectiles ($v\leq \SI{1}{\atomicunit}$) required smaller time steps of \SI{0.155}{\atto\second} (\SI{0.31}{\atto\second}) to converge AE (PAW) stopping powers in carbon.

For alpha-particle and proton projectiles, we included the corresponding ion in the static Mermin-DFT calculation for the initial condition.
All-electron calculations used a bare Coulomb potential for the projectile, which allowed for artificial, fractionally-charged projectiles.
To ensure a physical initial state, fractionally-charged projectiles were excluded from the Mermin-DFT calculation and suddenly inserted at the start of each TDDFT calculation.
This difference in initial condition only influences stopping forces during a short transient regime \cite{kononov2024reproducibility} that we ignored in the stopping power analysis.

To compute average stopping powers, we first calculated the stopping work
\begin{equation}
    W(x) = -\int_{x_0}^{x_f} \vec{F}(\vec{x}) \cdot d\vec{x} = -\int_{x_0/v}^{x_f/v} dt \vec{F}(t) \cdot \vec{v}, 
\end{equation}
where $\vec{x}$ is the projectile's displacement and $\vec{F}$ is the instantaneous force along its trajectory.
We then performed a linear fit and took the slope as the average stopping power:
\begin{equation}
    W(x) \approx S(x_f) x + W_0 \quad \text{for} \quad x_0<x<x_f.
    \label{eq:Sfit}
\end{equation}
We found that $x_0=\SI{4}{\angstrom}$ sufficed to exclude initial transient behavior.
Meanwhile, $x_f=\SI{80}{\angstrom}$ sufficed to representatively sample close collisions with host ions \cite{kononov2023trajectory} and converge average stopping powers in the solid-density aluminum and \SI{10}{\gram\per\centi\meter\cubed} carbon cases.
The \SI{1}{\gram\per\centi\meter\cubed} carbon case required a somewhat longer $x_f$ of \SI{120}{\angstrom}.

For very small test projectile charges ($Z\leq 0.1$), the stopping forces start to become dominated by extraneous contributions from the initial, unperturbed electron density, denoted $\vec{F}^{(0)}(\vec{x})$.
In principle, $\vec{F}^{(0)}$ should average to 0 over a sufficiently long trajectory so that its contribution to the average stopping power vanishes.
In practice, density inhomogeneities can introduce significant sensitivities to $x_f$ \cite{kononov2024reproducibility}.
To mitigate this numerical problem, we isolated the forces induced by electronic excitations, i.e., the force contributions relevant to average electronic stopping power.
To that end, we computed $\vec{F}^{(0)}(\vec{x})$ through a separate simulation where the projectile moves along its trajectory but the initial electron density remains fixed.
We then performed the fitting procedure using $\vec{F}-\vec{F}^{(0)}$ for the fractionally charged test projectiles.

\section{Results and Discussion}
\label{sec:results}

\begin{figure}
    \centering
	\includegraphics{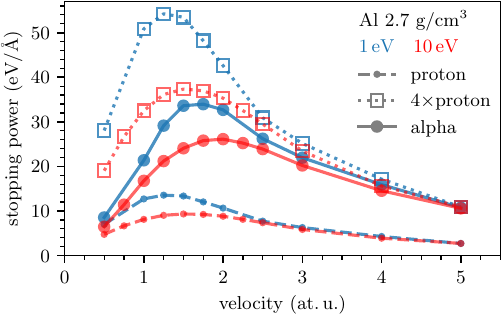}
    \caption{
    Free-electron contributions to stopping powers of protons (small circles) and alpha particles (large circles) in solid-density aluminum at an electronic temperature of \SI{1}{\electronvolt} (blue) and \SI{10}{\electronvolt} (red) computed with TDDFT.
    Empty squares indicate the proton data scaled by a factor of 4.
    }
    \label{fig:Alstopping}
\end{figure}

To demonstrate nonlinear stopping power effects and the need to account for the effective projectile charge in more efficient stopping power models, we first compare proton and alpha-particle stopping powers for the case of isochorically heated aluminum in Fig.~\ref{fig:Alstopping}.
At high velocities, where the projectile is expected to remain nearly fully ionized, the alpha-particle stopping powers are indeed about 4 times larger than the proton stopping powers.
However, near and below the Bragg peak, applying naive $Z^2$ scaling to the proton results would significantly overestimate the alpha-particle stopping powers.
In fact, at low velocities, alpha-particle stopping appears to converge toward the proton values, suggesting comparable effective charge states for both projectiles.
Furthermore, while the position of the stopping peak is typically considered a property of the host material alone --- determined by the dielectric function within linear-response theory (see Eq.~\eqref{eq:lr}) --- we find that it occurs at a slightly higher velocity for alpha particles than proton projectiles.

\begin{figure}
    \centering
    \includegraphics{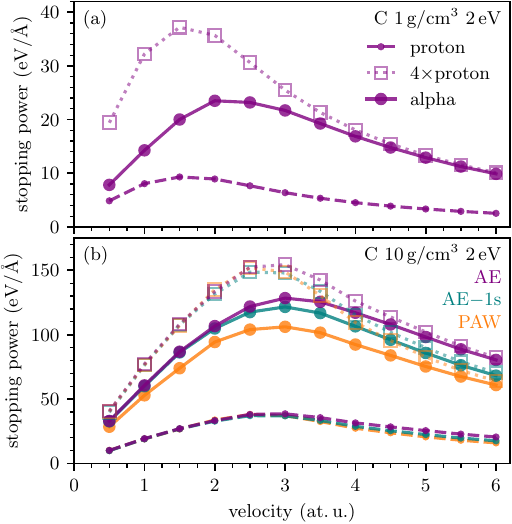}
    \caption{
    Stopping powers of protons (small circles) and alpha particles (large circles) in warm dense carbon at 
    a) \SI{1}{\gpcc}, \SI{2}{\electronvolt} and 
    b) \SI{10}{\gpcc}, \SI{2}{\electronvolt} computed with TDDFT.
    Results from all-electron (AE) calculations, all-electron calculations where carbon 1s states were frozen (AE$-$1s), and PAW calculations are shown in purple, teal, and orange, respectively.
    Empty squares show the proton data scaled by a factor of 4.
    }
    \label{fig:Cstopping}
\end{figure}

Although we find qualitatively similar behavior for light-ion stopping powers in carbon at \SI{1}{\gpcc} and \SI{2}{\electronvolt} (see Fig.~\ref{fig:Cstopping}a), the deviations from naive $Z^2$ scaling are more modest at \SI{10}{\gpcc} (see Fig.~\ref{fig:Cstopping}b).
The higher-density carbon case maintains significantly larger stopping powers for alpha particles compared to protons even at low velocities, contrasting with the behavior predicted for aluminum and lower-density carbon.
These qualitative differences can be understood in terms of the average electron densities: while the aluminum and \SI{1}{\gpcc} carbon cases have similar electron densities of \SI{0.027}{\atomicunit} and \SI{0.030}{\atomicunit}, the \SI{10}{\gpcc} carbon case has a significantly higher electron density of \SI{0.3}{\atomicunit}.
Thus, the free electrons in \SI{10}{\gpcc} carbon have more energy and are more resistant to capture, leaving the projectiles more ionized such that the stopping power follows $Z^2$ scaling more closely.
Notably, the absolute difference between the scaled proton stopping powers and the alpha-particle stopping powers is comparable in magnitude for both carbon densities.

The dense carbon system also allows us to efficiently assess the influence of pseudopotentials and core states.
We find the same trends for both all-electron (AE) calculations using bare Coulomb potentials and PAW calculations that pseudize the carbon 1s cores (see Fig.~\ref{fig:Cstopping}b).
For protons, the AE stopping powers modestly exceed the PAW values at high velocities mainly because of contributions from carbon 1s excitations.
Notably, the alpha-particle stopping powers are more sensitive to pseudization even at lower velocities, likely because the He PAW pseudopotential is somewhat softer than the H PAW pseudopotential.
The comparison between AE calculations with frozen carbon 1s states and PAW results suggests that projectile pseudopotential effects may cause around a 15\% underestimation of the free-electron contribution to alpha-particle stopping powers in aluminum.~
\footnote{A more accurate treatment of interactions between the projectile and free electrons is not practical for the aluminum case because VASP requires the same type of potential for all species, a harder He PAW pseudopotential is not readily available, and the strongly bound core states would require an extremely high cutoff energy in an AE aluminum calculation.}

\begin{figure}
    \centering
    \includegraphics{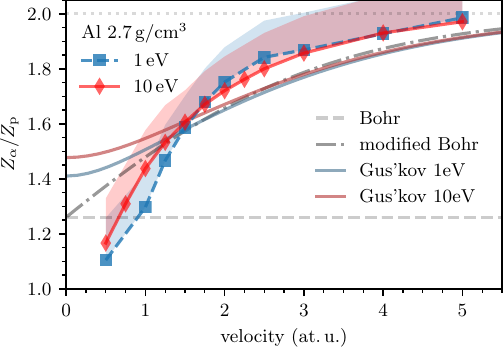}
    \caption{Ratio of effective charges $Z_\alpha/Z_\mathrm{p}$ in warm dense aluminum as inferred from TDDFT stopping powers according to Eq.~\eqref{eq:Zratio} at temperatures of \SI{1}{\electronvolt} (blue squares) and \SI{10}{\electronvolt} (red diamonds).
    Shading indicates estimated errors from approximated electron-ion potentials and time step convergence.
    Also shown are analytic predictions from the Bohr stripping criterion \cite{bohr:1948} (\mbox{Eq.~\eqref{eq:bohr}}), its modified version (Eq.~\eqref{eq:bohrmod}), and the condition-dependent generalization by Gus'kov et al.\cite{gus2009method} (Eq.~\eqref{eq:guskov}).
    Dotted gray indicates the high-velocity limit of $Z_\alpha/Z_\mathrm{p}\rightarrow 2$.
    }
    \label{fig:AleffectiveZ}
\end{figure}

\begin{figure}
    \centering
    \includegraphics{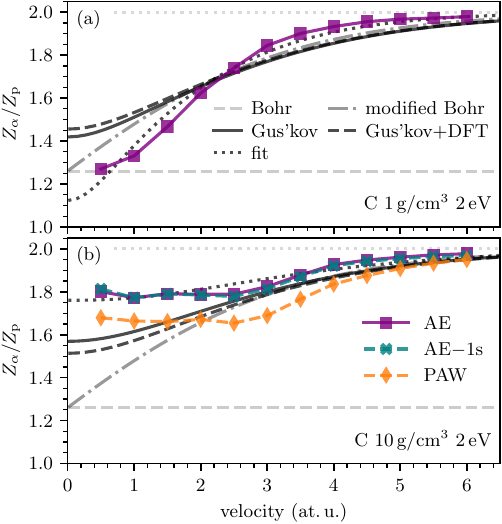}
    \caption{Ratio of effective charges $Z_\alpha/Z_\mathrm{p}$ in warm dense carbon at 
    a) \SI{1}{\gpcc}, \SI{2}{\electronvolt} and 
    b) \SI{10}{\gpcc}, \SI{2}{\electronvolt} as inferred from TDDFT stopping power calculations that include all electrons (purple squares), freeze carbon 1s states (teal exes), or use PAW to treat only valence electrons (orange diamonds).
    Shades of gray indicate analytic predictions from the Bohr stripping criterion \cite{bohr:1948} (Eq.~\eqref{eq:bohr}), its modified version (Eq.~\eqref{eq:bohrmod}), the condition-dependent generalization by Gus'kov et al.\cite{gus2009method} (Eq.~\eqref{eq:guskov}), the Gus'kov model with input from DFT (Eq.~\eqref{eq:vrelmod}), and a fitted version of the Gus'kov model (Eq.~\eqref{eq:guskovmod}).
    The high-velocity limit of $Z_\alpha/Z_\mathrm{p}\rightarrow 2$ is shown in dotted gray.
    }
    \label{fig:CeffectiveZ}
\end{figure}

Next, we evaluate the consistency of different effective charge models with the TDDFT data of Figs.~\ref{fig:Alstopping} and \ref{fig:Cstopping}.
If we assume that $S \propto Z_\mathrm{eff}^2$, then we can compute the velocity-dependent ratio of alpha-particle and proton effective charges as
\begin{equation}
    \frac{Z_\alpha(v)}{Z_\mathrm{p}(v)} = \sqrt{ \frac{S_\alpha(v)}{S_\mathrm{p}(v)} },
    \label{eq:Zratio}
\end{equation}
where $S_\alpha$ and $S_\mathrm{p}$ denote alpha-particle and proton stopping powers computed using TDDFT.
According to the original Bohr stripping criterion (Eq.~\eqref{eq:bohr}), this effective charge ratio would have a constant value of $2^{1/3}\approx 1.26$.
Meanwhile, the modified models of Eqs.~\eqref{eq:bohrmod} and \eqref{eq:guskov} predict velocity-dependent ratios that correctly asymptote to the ratio of nuclear charges in the high-velocity limit.
We compare the effective charge ratios predicted by each analytic model described in Section~\ref{sec:models} to values inferred from the TDDFT stopping power data in Figs.~\ref{fig:AleffectiveZ} and \ref{fig:CeffectiveZ}.
We generally find satisfactory agreement for ion velocities near and above the Bragg peak, but discrepancies at low velocities may warrant further scrutiny.

In solid-density aluminum, TDDFT predicts a significantly steeper velocity dependence for the effective charge ratio of slow ions with $v\lesssim\SI{2}{\atomicunit}$ than any of the analytic models (see Fig.~\ref{fig:AleffectiveZ}).
Within TDDFT, the low-velocity limit of $Z_\alpha/Z_\mathrm{p}$ appears to approach 1 rather than the Bohr value of 1.26 or the Gus'kov value of 1.41\,--\,1.48.
While the Bohr value falls within the uncertainties of the TDDFT data (given by the product of estimated relative errors from time-step convergence and pseudopotential effects), the discrepancy with Gus'kov's low-velocity limit cannot be explained by the sources of errors considered in this work.
Nonetheless, the temperature-dependence of the Gus'kov model matches the trend in the TDDFT data, with a higher effective charge ratio at a temperature of \SI{10}{\electronvolt} than \SI{1}{\electronvolt}.

In the low-density carbon case of Fig.~\ref{fig:CeffectiveZ}a, $Z_\alpha/Z_\mathrm{p}$ exhibits similar qualitative behavior as in solid-density aluminum. 
The low-velocity limit predicted by TDDFT in \SI{1}{\gpcc} carbon recovers the Bohr stripping criterion, suggesting that this case sufficiently resembles the cold gas limit considered by Bohr.
In contrast, the TDDFT results for the high-density carbon case of Fig.~\ref{fig:CeffectiveZ}b maintain a relatively large effective charge ratio even for slow ions.
Although the Gus'kov model also predicts that $Z_\alpha/Z_\mathrm{p}$ increases with density, the magnitude of this effect is much larger within TDDFT.

While the uniform electron gas assumed by the Gus'kov model represents aluminum fairly well, free electrons in warm dense carbon are not as ideal.
At both carbon densities considered, the DFT density of states contains a sharp feature near the onset of the continuum that modifies the velocity distribution of occupied states.
To take this nonideality into account, we modify the Gus'kov model by evaluating $v^*$, the root-mean-square projectile velocity relative to host electrons, using information from DFT.
That is, we take
\begin{align}\begin{split}
    & (v^*)^2 = v^2 + \langle v_\mathrm{e}^2\rangle, \\
    & \langle v_\mathrm{e}^2\rangle = \frac{2}{n_e} \int_0^\infty dE \; (E-E_0) \, f(E) \, \mathrm{DOS}(E),
    \label{eq:vrelmod}
\end{split}\end{align}
where $\langle v_\mathrm{e}^2\rangle$ is the mean-square velocity of free electrons, $E-E_0$ is the energy relative to the continuum onset, $f(E)$ is the Fermi distribution, and $\mathrm{DOS}(E)$ is the density of states per unit volume as predicted by DFT.
Although the DFT estimates of $\langle v_\mathrm{e}^2\rangle$ in carbon differ from the uniform electron values used in Eq.~\eqref{eq:guskov} by a factor of about 1.6, this modification has only a minor effect on the effective charge predictions of Fig.~\ref{fig:CeffectiveZ}.

The Gus'kov model also contains empirical choices describing the transition between the low-velocity Bohr-like limit and the high-velocity bare ion charge.
In an attempt to improve these choices, we perform least-squares fits using a parameterized form,
\begin{equation}
    Z'_\text{Gus'kov} = Z\left( 1 - p_1 e^{-p_2 v^* / Z^{2/3} }\right)
    \label{eq:guskovmod},
\end{equation}
where $p_1, p_2$ are fit parameters.
For aluminum, we use the analytic expression for $v^*$ given in Eq.~\eqref{eq:guskov}, while for carbon we use the $v^*$ values given by the DFT density of states (see Eq.~\eqref{eq:vrelmod}).
To take into account the estimated errors in the aluminum data (see shading in Fig.~\ref{fig:AleffectiveZ}), we take the union of parameter ranges obtained from two fits, one using the computed TDDFT data and the other including estimated corrections.
Although the additional flexibility of the parameterized form improves the model's ability to capture the TDDFT results (see Fig.~\ref{fig:CeffectiveZ}), the best fit parameters depend on the material and conditions (see Table \ref{tab:Zeff_fit}).
Thus, a universal effective charge model remains elusive.

\begin{table*}
    \centering
    \begin{tabular}{c|c|c|c|c|c}
            & Gus'kov \cite{gus2009method} & C \SI{1}{\gpcc} \SI{2}{\electronvolt} & C \SI{10}{\gpcc} \SI{2}{\electronvolt} & Al \SI{2.7}{\gpcc} \SI{1}{\electronvolt} & Al \SI{2.7}{\gpcc} \SI{10}{\electronvolt} \\\hline
        $p_1$ & 1     & $1.48\pm 0.07$  &  $0.58\pm 0.04$ & $1.95\pm 0.34$ & $2.56\pm 0.68$ \\
        $p_2$ & 0.92  & $1.28\pm 0.07$  &  $0.82\pm 0.07$ & $2.08\pm 0.49$ & $2.17\pm 0.54$ \\
        $\mathrm{corr}(p_1,p_2)$ & --- & 0.90 & 0.74 & 0.93\,--\,0.94 & 0.95\,--\,0.97
    \end{tabular}
    \caption{Effective charge model parameters $p_1$ and $p_2$ (see Eq.~\eqref{eq:guskovmod}) as proposed by Gus'kov et al.~\cite{gus2009method} and obtained from least-squares fits to the TDDFT data shown in Figs.~\ref{fig:AleffectiveZ} and \ref{fig:CeffectiveZ}.
    Parameter uncertainties were determined from the diagonal elements of the fit covariance matrix and the estimated TDDFT errors in the aluminum case (see text).
    The last row gives the correlation coefficient between the fit parameters.
    }
    \label{tab:Zeff_fit}
\end{table*}

To benchmark the effective charge models more directly, we assess nonlinear effects in the \SI{10}{\gram\per\centi\meter\cubed} carbon case using fractionally charged test projectiles.
Just like the proton- and alpha-particles in the all-electron carbon calculations, these test projectiles are represented by a moving Coulomb potential $-Z/|\mathbf{r}-\mathbf{v}t|$, where $Z$ ranges from 0.01 to 2.5.
Sufficiently small $Z$ would be expected to recover the linear-response regime, where changes in the electron density become proportional to the perturbation strength and the stopping power scales as $Z^2$.
Thus, we estimate effective charges as
\begin{equation}
    Z_\mathrm{eff}(Z,v) = \sqrt{S_Z(v)} \lim_{Z'\rightarrow 0} \frac{Z'}{\sqrt{S_{Z'}(v)}},
\end{equation}
where $S_Z(v)$ are computed with TDDFT and the value of the $Z'\rightarrow 0$ limit relates to the dielectric integrals of Eq.~\eqref{eq:lr}.
We focus on two contrasting velocities: $v=2$ and \SI{5}{\atomicunit}, where $Z_\alpha/Z_\mathrm{p}$ lies near its low-velocity and high-velocity asymptotic values, respectively (see Fig.~\ref{fig:CeffectiveZ}b).

\begin{figure}
    \centering
    \includegraphics{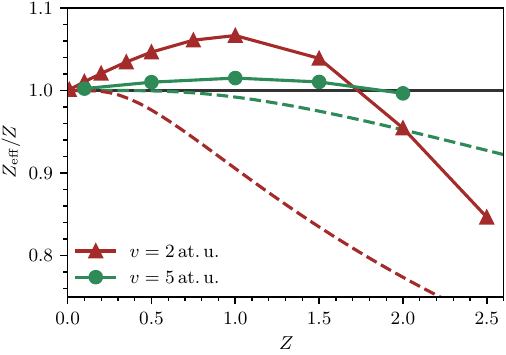}
    \caption{Effective ionization of test charges stopping in carbon at \SI{10}{\gram\per\centi\meter\cubed} and \SI{2}{\electronvolt}.
    Brown triangles and green circles represent AE TDDFT results at two different velocities, while corresponding dashed curves indicate behavior predicted by the Gus'kov effective charge model \cite{gus2009method}.
    The horizontal line indicates $Z_\mathrm{eff}=Z$.
    }
    \label{fig:CfractionalZ}
\end{figure}

Contrary to common assumptions, we find that proton stopping power already exhibits some nonlinear behavior with $Z_\mathrm{eff}$ deviating noticeably from $Z$, particularly at the lower velocity of \SI{2}{\atomicunit} (see Fig.~\ref{fig:CfractionalZ}).
The Gus'kov effective charges generally fall below the TDDFT results, and the functional forms of the analytic models considered in Section~\ref{sec:models} are not capable of describing the non-monotonic behavior of the TDDFT effective charges.
Nonetheless, better qualitative agreement could be possible for faster and/or higher $Z$ projectiles.

The $Z_\mathrm{eff}>Z$ predictions in Fig.~\ref{fig:CfractionalZ} suggest that the linear response assumption breaks down for velocities near the Bragg peak.
In this regime, contributions from higher-order response functions only become negligible for projectiles with unphysically low $Z$, and they cannot be approximated by a physically reasonable effective charge.
The apparent shape of $Z_\mathrm{eff}/Z$ in Fig.~\ref{fig:CfractionalZ} can be attributed to additional stopping power contributions with $Z^3$ and $Z^4$ scaling \cite{ashley1972z, lindhard1976barkas}, known as the Barkas-Andersen effect \cite{barkas1963resolution,andersen1977stopping} and Bloch correction \cite{bloch1933bremsung}, respectively.

Finally, we compare the linear-response limit predicted by TDDFT to Lindhard's linear-response formula (Eq.~\eqref{eq:lr}), effectively removing the complications of nonlinear effects for isolated benchmarking of dielectric models.
Here, we focus on free-electron contributions, i.e., the carbon 1s states remained frozen within the TDDFT simulations.
The integrals in Eq.~\eqref{eq:lr} were evaluated as described in Refs.~\onlinecite{hentschel:2023,hentschel:uegcode} using dielectric functions given by the random phase approximation (RPA) or the Mermin approximation \cite{mermin:1970}, which captures electron-ion collisions beyond RPA.
The collision frequencies informing the Mermin dielectric function were obtained from a T-matrix formulation within an average-atom model \cite{hentschel:2023}.

\begin{figure}
    \centering
    \includegraphics{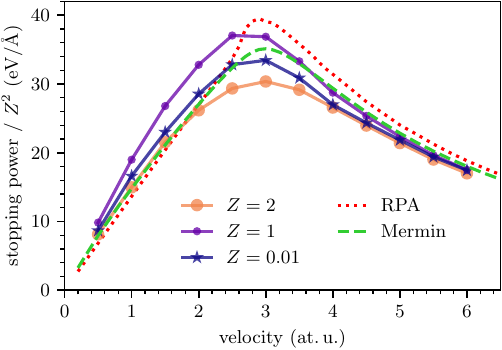}
    \caption{Comparison between free-electron stopping powers obtained with TDDFT and linear-response theory for carbon at \SI{10}{\gram\per\centi\meter\cubed} and \SI{2}{\electronvolt}.
    Circles indicate experimentally relevant TDDFT predictions for alpha particles and protons, while stars represent the theoretical linear-response limit computed with TDDFT through a fictitious $Z=0.01$ projectile.
    Dotted red and dashed green curves show linear-response results evaluated according to Eq.~\eqref{eq:lr} using RPA and Mermin model dielectric functions, respectively.
    }
    \label{fig:Cdielectric}
\end{figure}

The Mermin dielectric framework agrees quite well with TDDFT stopping powers, particularly in the low-Z limit approaching the linear response regime (see Fig.~\ref{fig:Cdielectric}).
Thus, the modest discrepancies between TDDFT and Mermin stopping power predictions for protons and alpha particles mainly arise from the nonlinear behavior captured in the TDDFT calculations, with discrepancies in dynamic response functions \cite{hentschel:2023,hentschel2025statistical} largely canceling out after energy and momentum integration.
These findings suggest that further improvements to efficient stopping power models will require including contributions beyond linear response.

\section{Conclusions}
\label{sec:conclusions}

Our real-time TDDFT predictions reveal nonlinear effects in light-ion stopping powers, particularly at velocities near and below the Bragg peak where $v$ is comparable to or less than $Z^{2/3}$.
While efficient stopping models typically assume linear response, where stopping power scales quadratically with the projectile's charge, we find that actually approaching the linear-response regime can require an unphysically small test charge.
In particular, even proton stopping powers appear to contain noticeable contributions from nonlinear behavior.

Effective charge models could approximately capture some qualitative aspects of our TDDFT results by replacing the bare nuclear charge with an appropriately screened ion charge.
The most sophisticated effective charge model considered in this work \cite{gus2009method} predicts variations with velocity, temperature, and density that are largely consistent with qualitative trends in the TDDFT data.
However, other types of nonlinear effects (besides partial projectile neutralization) preclude quantitative agreement at low velocities.

These findings will inform ongoing efforts to benchmark and improve efficient stopping power models based on accurate TDDFT calculations \cite{hentschel:2023,stanek_review_2024,nichols2023time,white2022mixed}.
Since earlier work on light-ion stopping powers typically considered nonlinear effects negligible, discrepancies between TDDFT data and linear-response treatments were primarily attributed to deficiencies in the model dielectric function used within the latter approach \cite{hentschel:2023,white2022mixed}.
We demonstrated that computing TDDFT stopping powers for small test charges can isolate linear-response contributions and provide more focused benchmark data.
Accurately capturing nonlinear effects within computatonally efficient stopping power models remains a topic for future research.

\begin{acknowledgements}
We are grateful to the organizers of the 2nd Charged-Particle Transport Coefficient
Comparison Workshop --- including Lucas J. Stanek, Brian M. Haines, Patrick F. Knapp, Michael S. Murillo, Liam G. Stanton, and Heather D. Whitley --- for catalyzing this work.
We also thank 
Alexander White, Nathaniel Shaffer, Paul Grabowski, and Alexandra Olmstead for fruitful discussions, 
Cody Melton for graciously providing thermalized carbon configurations,
Joel Stevenson for HPC support,
and Heath Hanshaw for pre-publication review.
All authors were partially supported by the US Department of Energy Science Campaign 1 and Sandia National Laboratories' Laboratory Directed Research and Development (LDRD) Project Nos.\ 218456 and 233196.

This work was performed, in part, at the Center for Integrated Nanotechnologies, an Office of Science User Facility operated for the U.S.\ Department of Energy (DOE) Office of Science.
Sandia National Laboratories is a multi-mission laboratory managed and operated by National Technology \& Engineering Solutions of Sandia, LLC (NTESS), a wholly owned subsidiary of Honeywell International Inc., for the U.S. Department of Energy’s National Nuclear Security Administration (DOE/NNSA) under contract DE-NA0003525. This written work is authored by an employee of NTESS. The employee, not NTESS, owns the right, title and interest in and to the written work and is responsible for its contents. Any subjective views or opinions that might be expressed in the written work do not necessarily represent the views of the U.S.\ Government. The publisher acknowledges that the U.S.\ Government retains a non-exclusive, paid-up, irrevocable, world-wide license to publish or reproduce the published form of this written work or allow others to do so, for U.S.\ Government purposes. The DOE will provide public access to results of federally sponsored research in accordance with the DOE Public Access Plan.
\end{acknowledgements}

\bibliography{main.bib}

\end{document}